# A New Strategy For Solving Two Cosmological Constant Problems In Hadron Physics

**Thomas L. Wilson**
NASA, Johnson Space Center - KR, Houston, Texas USA
Email: Thomas.Wilson@cern.ch

February 21, 2013

**Abstract**

A new approach to solving two of the cosmological constant problems (CCPs) is proposed by introducing the Abbott-Deser (AD) method for defining Killing charges in asymptotic de Sitter space as the only consistent means for defining the ground-state vacuum for the CCP. That granted, Einstein gravity will also need to be modified at short-distance nuclear scales, using instead a nonminimally coupled scalar-tensor theory of gravitation that provides for the existence of QCD's two-phase vacuum having two different zero-point energy states as a function of temperature. Einstein gravity alone cannot accomplish this. The scalar field will be taken from bag theory in hadron physics, and the origin of the bag constant $B$ is accounted for by gravity's CC as $\lambda_{Bag} = \bar{\kappa} B$ – noting that the Higgs mechanism does not account for either the curved-space origin of $\lambda$ or the mass of composite hadrons. A small Hubble-scale graviton mass $m_g \sim 10^{-33} eV$ naturally appears external to the hadron bag, induced by $\lambda \neq 0$. This mass is unobservable and gravitationally gauge-dependent. It is shown to be related to the cosmological event horizon in asymptotic de Sitter space.

## 1. Introduction

The cosmological constant problem (CCP) continues to represent a serious circumstance for the unification of gravity with quantum field theory (QFT) on curved backgrounds, quantum gravity (QG), zero-point energy fluctuations, and our understanding of vacuum energy density $\rho_V$ in particle physics as well as cosmology. The point of view to be taken here is that the CCP(s) cannot be fixed without two things: (a) a consistent definition and usage of global energy (Killing charge) in asymptotic spacetime; and (b) a modification of Einstein gravity (EG) where the cosmological constant (CC) $\lambda$ was first introduced and discovered [1,2]. It has been subsequently identified as a vacuum energy density (VED) [3-5] which is a gravitational effect resulting in a curved de Sitter spacetime referred to as cosmological gravity (CG) (metrics with $\lambda \neq 0$). The dilemma presented by various proposed CCP's today is the disparity between cosmological measurements of $\rho_V$ [6-8] and those calculated in particle [9] and hadron physics.[1]

In parallel with EG, relativistic QFT has pursued VED physics in flat Minkowski space, resulting for example in the remarkable spontaneous symmetry breaking (SSB) mechanism that will be used later in Sect. 3. Even though EG is nonrenormalizable, its gravitational field $g_{\mu\nu}$ couples minimally and universally to all of the fields of QFT's renormalizable standard model [10]. To turn on gravity one simply introduces EG along with covariant derivatives in QFT that represent the transition from flat to curved background metrics. This ties together everything except for one major shortcoming, the gravitational versus flat-space VED problem usually referred to as the CCP. Hence there are dramatic differences in QFT and its renormalization when cosmological gravity becomes involved.[2]

The literature [11,12] speaks of using Minkowski counterterms to subtract from the bare $\rho_V$ in Friedmann-Lemaitre-Robertson-Walker (FLRW) cosmology in order to derive the physical, renormalized VED $\rho_{ren}$ in curved spacetime. This creates major confusion. It uses flat metrics to fix the asymptotic states of cosmological metrics that are never flat. One theme here will be to take this reasoning a step further to preclude cross-comparison of asymptotic spacetimes. This will involve the Abbott-Deser (AD) method of identifying mass and energy [13] and their Killing-charge successors as the unique quantities associated with the asymptotic geometry at spatial infinity of de Sitter spacetime.

---

[1] Note that this study does not address "the" CCP because it is becoming increasingly apparent that we still do not seem to understand what "the CCP" is. Instead, the approach here is to try and define two CCPs, and devise a consistent method for solving them later when satisfactory theories of QG and QCD (quantum chromodynamics) confinement exist.

[2] Double-counting is commonplace in current methods, introducing $\rho_V$ in both (9) and (11)-(14) that follow. This affects the renormalization loop equations. An attempt to preclude double-counting will be made here.

As mentioned, it has been said that there are at least two CCP's [14,15]. (a) The old one is to understand why $\rho_V$ measured by current Type Ia supernova observations [6-8],

$$\lambda < 10^{-56} \, cm^{-2} \sim 10^{-29} \, g \, cm^{-3} \sim 10^{-47} \, GeV^4, \quad (1)$$

(where $\lambda$ is positive) is orders of magnitude smaller than values suggested by particle physics where $\rho_{QCD} \sim 10^{-6} \, GeV^4$ appears in QCD [9] or the bag constant $B \sim (145 \, MeV)^4$ in hadron physics. This is the fine-tuning problem of reconciling (1) with expectations from particle physics. It also is illustrated by the cutoff $\Lambda_c$ in effective field theory represented by the Planck mass as $<\rho_{Pl}> \sim \Lambda_c^4 \sim 10^{71} GeV^4$.[3] (b) In contrast, the second CCP [14,15] is to understand why $\rho_\lambda$ is of the same order of magnitude as the present mass density of the universe $\rho_M$ in Freidmann-Lemaitre-Robertson-Walker (FLRW) cosmology.

We will paraphrase the CCP(s) differently here with the goal of identifying a well-posed statement of two of the problems. Then an attempt will be made to show how they can be approached in hadron physics using scalar-tensor theory.

The original CCP assumed here (CCP-1) is to explain how gravity theory can have two different vacuum energy states or two different cosmological constant (CC) values when Einstein gravity only permits one as a universal constant throughout all of spacetime. The second (CCP-2) is to understand how quantum fluctuations relate to the ground-state energy of curved spacetime in order to define the zero-point energy of the gravitational background in a consistent fashion. Resolving CCP-2 is essential to defining the ground-state energy for investigating CCP-1.

We will take these in reverse order, addressing the zero-point fluctuation issue in FLRW cosmology in Sect. 2, and developing the scalar-tensor model in Sec. 3. Experimental questions will be addressed in Sect. 4. Assumptions and postulates are identified in Sect. 5. And conclusions follow in Sect. 6. An Appendix gives the derivation of the cosmological constant as a gauge-dependent graviton mass $m_g$ in the weak-field limit, a result that is directly related to the AD formalism and the cosmological event horizon present in asymptotic de-Sitter spacetime and the FLRW universe (Sect. 4). At late times the latter presently behaves like an accelerating de-Sitter spacetime [6-8], discussed in Sect. 2.

The appearance of a graviton mass [see derivation in Appendix, (65)-(66)] is natural, following directly from EG with $\lambda \neq 0$ and having an obvious smooth zero-mass limit $m_g \to 0$ as $\lambda \to 0$. It is not introduced *ad hoc* in the usual manner based upon adding the Pauli-Fierz Lagrangian for a mass $m_g = m_{PF}$ (where $\lambda=0$ was originally assumed) adopted in particle physics.

## 2. Zero-Point Vacuum Fluctuations

### 2.1. A Digression on Asymptotic de Sitter Space

It is well known that the Schwarzschild-de-Sitter metric (SdS) [16]

$$ds^2 = -c(r)dt^2 + c^{-1}(r)dr^2 + r^2 d\Omega^2, \quad (2)$$

where

$$c(r) = 1 - \frac{2m}{r} - \frac{\lambda}{3}r^2, \quad (3)$$

represents important global properties that relate to the definition of energy and energy conservation in Einstein theory. In (2) and (3), we have $m = GM/c^2$ with $d\Omega^2 = (d\theta^2 + \sin\theta d\varphi^2)$ a unit 2-sphere metric, and $M$ the Schwarzschild mass.[4]

Arnowitt, Deser, and Misner (ADM) [17] were first to derive a canonical formulation of general relativity (GR) as a Hamiltonian system for the simple Schwarzschild case ($\lambda=0$) in (3). They determined the ADM energy, momentum, and mass defined by the asymptotic symmetries of (2) and (3) at spatial infinity. Conserved charge (mass, energy, etc.) is associated with a conserved vector (Noether) current which is determined by reducing the stress tensor density conservation law $\nabla_\mu T^{\mu\nu}$ in EG to a conserved vector current law using Killing vectors $\xi^\mu$. The ADM mass results and is equivalent to the Schwarzschild mass $M$, $M_{ADM}=M$ in (3).

Note that the Schwarzschild metric, (2) and (3) with ($\lambda=0$), is asymptotically flat as $r\to\infty$. Note also that assuming $M=0$ and $\lambda=0$ results in flat Minkowski space. In either case, the energy of Minkowski space is zero as expected. Classically speaking, it likewise has no VED. In this sense, the natural vacuum of EG without $\lambda$ is flat space with all its Poincaré symmetries.

Re-instating $\lambda \neq 0$ in (3), however, changes circumstances significantly. The full SdS metric (2) is not asymptotically flat and becomes an asymptotic de Sitter space as $r\to\infty$ that is forever distinguished from Min-kowski space.[5] When a CC $\lambda$ is present, flat Minkowski space is no longer a relevant background because it is not a solution of the Einstein equations [13]. The vacuum is now either de Sitter [SO(4,1)] or anti-de Sitter [SO(3,2)] depending upon whether $\lambda$ is positive or negative.[6]

---

[3] Note that the quartic cut-offs $\Lambda_c^4$ are QFT values derived in flat Minkowski space. These imply a gravitational curvature $10^{118}$ times that in (1) in EG where the Ricci scalar $R = 4\lambda$ is not flat. Curvature and EG are ignored completely. This is the inconsistent cross-comparison problem that will be addressed in Sect. 2.

[4] In general, natural units $\hbar = c = 1$, metric signature $(-,+,+,+)$, and a 4-dimensional spacetime are assumed.

Abbott and Deser [13,19] extended the fundamental ADM approach used in the Schwarzschild case and defined the AD Killing charges for the full SdS metric when it asymptotically becomes de Sitter space (dS), as opposed to the asymptotic flat case above. These AD charges have become very important because of their direct relevance to cosmological gravity and, as will be shown here, the CCP. This work has been extended by Deser and Tekin [20-23], and the collective results will be referred to as the ADT formalism.

There is an apparent singularity in (3) for $m = 0$ at $r_{EH} = \sqrt{3/\lambda}$, that keeps the observer from proceeding smoothly to infinity. Gibbons & Hawking (G-H) [24] developed $r_{EH}$ as a cosmological event horizon characterizing asymptotic dS whose surface gravity is $\kappa_c = r_{EH}^{-1}$. AD [13] further pointed out that the Killing vector $\xi^\mu$ is timelike only within the background cosmological event horizon $r < r_{EH}$. The usual meaning of global energy $E$ and the timelike Killing vector are lost on the super-horizon scale for $r > r_{EH}$. This will be discussed further in Sect. 4.

Adding Weyl and Gauss-Bonnet quadratic curvature terms[7] (scaled by $\alpha$ and $\beta$ respectively) to the Einstein-Hilbert Lagrangian [20,21], Deser and Tekin generalized the AD mass to

$$E = M + 8\lambda M \kappa_{GB}^{-1}(4\alpha + \beta) + \kappa_{EG}^{-1}\lambda V_{dS}, \quad (4)$$

where the term $V_{dS}$ is the volume of the spacetime and has been added here to account for the pure dS case with $M=0$ in (3) and (4), involving higher-order terms not addressed by ADT. Dividing (4) by $V_{dS}$ to create an energy density, this same term has been found by Padmanabhan [29] using different methods.

Obviously, an empty dS with a VED due to $\kappa_{EG}^{-1}\lambda$ will contain a vacuum energy of that amount, $E \sim \kappa_{EG}^{-1}\lambda V_{dS}$. In an infinite dS space with finite VED, there is an infinite $E$. For a finite $V_{dS}$, then $E$ is finite such as within the G-H cosmological horizon $r_{EH}$.[8]

In summary, the total gravitational energy $E$ of spacetime is well-defined using ADM and ADT methods, provided it is being compared with a metric that has the same asymptotic structure [30]. However, there is a caveat. Comparison of energies between asymptotically flat Minkowski and asymptotically de Sitter metrics is a misleading exercise because the concepts of global energy and energy conservation become ill-defined in EG. Insistence upon comparison will result in an infinite energy between the two spacetimes.[9]

Cross-comparison of cosmological gravity with flat metrics contributes to the disparity in the old CCP where conclusions are being drawn based upon a comparison of incompatible asymptotic spacetime vacuum states in EG for dS versus flat QFT. Such comparison breaks the principle of compatible asymptotic states [9]. Yet this procedure is commonplace in the CCP literature, an example being the fine tuning problem where quadratic and quartic divergences in flat Minkowski space are being compared with asymptotically pure de Sitter (APdS) spacetime in cosmological gravity.

### 2.2. Ground State Vacua in Asymptotic de Sitter Space

First and foremost, we must recognize that FLRW cosmology is the basis for conclusion (1). The metric is

$$ds^2 = -dt^2 + a(t)^2 d\Omega_K^2, \quad (5)$$

where $a$ is the scale factor and

$$d\Omega_K^2 = \frac{(dr^2)}{(1-Kr^2)} + r^2 d\Omega^2, \quad (6)$$

with Gaussian curvature $K = 0$ [31-33]. In its late stages (current epoch), (5) is asymptotically an accelerating de Sitter space determined by the cosmological parameter $q = -\ddot{a}a/\dot{a}^2$ as derived from the Einstein-Friedmann equations [33].

The global energy of this universe is determined by the ADT charges for APdS spacetime with $M = 0$ in (2), (3), and (4) (no ADM mass). This is a critical point to make when defining the zero-point vacuum fluctuations and renormalization issues in general for the CCP.

The suggestion by Maggiore et al. [11,12] for re-defining the counter-term subtraction scheme to eliminate the quartic divergence $\Lambda_c^4$ is a promising idea. However, it is beset with at least one problem. It violates the compatibility principle of Sect. 2.1. Renormalization counter-term methods must conform with the ADT prescription for APdS spacetimes involving cosmological gravity having metric $g_{\mu\nu}$. Flat Minkowski space is not relevant because there is no $g_{\mu\nu}$ in that metric.

That is, the metric $g_{\mu\nu}$ of cosmological gravity typically is decomposed into a background $\eta_{\mu\nu}$ plus a fluctuation or perturbation $h_{\mu\nu}$ of arbitrary strength,

$$g_{\mu\nu} \equiv \eta_{\mu\nu} + h_{\mu\nu}, \quad (7)$$

---

[5] For a different interpretation, see [18].

[6] Negative $\lambda$ will not be considered, for reasons given later in Sect. 4.2.

[7] These have been characterized as improving the renormalizability of QG [25,26] although at the price of sacrificing unitarity [27,28].

[8] There is no coordinate invariant gravitational energy or energy density of a finite volume. The global energy of the total spacetime is well-defined but only with respect to another spacetime having the same asymptotic structure [30].

[9] This subject will be elevated to a principle of compatible asymptotic structure or states (Sect. 5).

where the background $\eta_{\mu\nu}$ is often defined symbolically as $\eta_{\mu\nu} \rightarrow \bar{g}_{\mu\nu}$. When $\eta_{\mu\nu} \rightarrow (-1,+1,+1,+1)$ the result is flat Minkowski space which has no gravity.

Following ADM and AD, the energy $E$ can be obtained from the Hamiltonian in cosmological gravity $H_{CG}$ by

$$E \equiv H_{CG}[g_{\mu\nu}] - H_{CG}[\eta_{\mu\nu}] \quad , \quad (8)$$

where $\eta_{\mu\nu}$ must be the APdS spacetime representing FLRW cosmology in the current epoch, an accelerating dS with $\lambda \neq 0$. That granted, assuming that $\eta_{\mu\nu}$ is flat Minkowski space violates the principle of compatibility. The ground-state vacuum of nonflat APdS spacetime has little to do with the ADM charges that derive from the asymptotically flat Schwarzschild metric and flat Minkowski spacetime.

The standard textbook procedure for analyzing quantum vacuum fluctuations inspired by inflation in cosmology is given by Weinberg [34] and is the same method that appears in the ADT procedure discussed above, while adopting (7) [13,20-23]. The gravitational field equations [below in (9)] are separated into a part linear in $h_{\mu\nu}$ plus all of the nonlinear terms that constitute the total source, the stress tensor $T_{\mu\nu}$ which is conserved using the Bianchi identities. Hence, global conservation of energy-momentum in the universe is assumed in these derivations. However, the catch like the caveat is that the global Killing charges may not be understood or consistently defined.

Having made the point that the origin of the CCP originates in (1), which derives from FLRW cosmology and which is currently interpreted as an accelerating de Sitter phase, the asymptotically pure de Sitter metric is the vacuum ground state for addressing this problem, as depicted in (8). This is not an assumption. It is the only conclusion that seems to follow from consistency and the principle of compatible asymptotic states (Sect. 2.1 and 5). Flat Minkowski spacetime is not relevant because it has no $g_{\mu\nu}$, breaks the principle, and invokes the caveat.

We are now prepared to advance to CCP-1 which in our opinion is the original and most important problem to address.

## 3. Modifying Einstein Gravity

Attempts to modify EG are nothing new. The real issue is the motivation for doing so. Einstein's theory[10]

$$R_{\mu\nu} - \frac{1}{2}g_{\mu\nu}R + \lambda g_{\mu\nu} = -\kappa T_{\mu\nu} \quad , \quad (9)$$

is remarkably successful on long-distance scales from binary pulsars [35] and planetary orbits [36] to short-distances of 1 mm [37].

---

[10] $R$ is the scalar curvature, $R_{\mu\nu}$ is the Ricci tensor, $g_{\mu\nu}$ is the spacetime metric, $T_{\mu\nu}$ is the energy-momentum tensor, and $\kappa = 8\pi G/c^4$ with $\bar{\kappa} = \kappa c^2$ where $G$ is Newton's gravitation constant, $c$ is the speed of light, and $x=(\mathbf{x},t)$.

However, one of the lessons from particle physics and QFT has been that SSB clearly involves a scalar field (below in Sect. 3.2) which generates a VED contribution to the CCP. Furthermore, SSB is involved in the bag model whose scalar has been proposed as responsible for confinement in hadron physics (below in Sect. 3.3) since there is no scalar in QCD [38] save for gluon and quark condensates.

Because hadrons comprise most of the matter in the universe, such a scalar field must be a gravitational one since only gravity is coupled universally to all physics. It couples attractively to all hadronic matter in proportion to mass and therefore behaves like gravitation similar to the scalar Spin-0 component of a graviton. Also, hadrons are a primary example of SSB known to exist experimentally and whose VED is determinable in the laboratory.

This means that a JFBD-type scalar-tensor theory of gravitation [39-41] should be an obvious candidate for modifying (9) in order to incorporate the SSB features of bag theory and hadron physics into gravitation theory at sub-mm scales. Einstein gravity has prevailed experimentally over JFBD scalar-tensor theory since the parameter $\Omega$ appearing in the latter has planetary time-delay measurements that place it at best as $\Omega \geq 500$ while Cassini data indicates it may be $\Omega \geq 40,000$ [42-44]. Therefore $\Omega$ is very large and JFBD $\rightarrow$ EG, although there are exceptions to this limit [45].

Hence EG and the Newtonian inverse-square law are the correct theory of gravity above 1 mm. The use of JFBD theory here will only introduce new experimental possibilities at sub-mm scales involving hadrons where $\Omega$ has never been measured. This modification will not change experimental EG as we currently understand it.

### 3.1. Breaking Lorentz Invariance

A remarkable property of (9) is its cosmological term $\lambda g_{\mu\nu}$, a fact that did not go unnoticed by Zel'dovich [5]. The energy level of the vacuum state (as in Sect. 2) must be defined. The first obvious point is that the Einstein vacuum in (9) is Lorentz invariant. Its stress tensor $T_{\mu\nu}$ must be the same in all frames. As a consequence, its vacuum average value can only be of the type like the Einstein term $\lambda g_{\mu\nu}$,

$$<0|T_{\mu\nu}|0> = \varepsilon_{vac} g_{\mu\nu} \quad . \quad (10)$$

That the QFT vacuum is Lorentz invariant as in (10) is a fundamental cornerstone of QCD [46]. Similarly, the vacuum fluctuations (again Sect. 2) in QCD have infinitely many degrees of freedom, contributing an infinite energy to (10). These are gotten rid of by "renormalization"; physicists are rarely interested in the very high-frequency modes, so their zero-point energy is assumed to be an unimportant additive constant which can be set to zero [46].

Now consider the APdS vacuum in (7). Physicists as observers can never "see" beyond the Gibbons-Hawking event horizon $r_{EH}$. But the global Killing charges in (4) are typically conserved, and this guarantees that the vacuum fluctuations on the APdS vacuum for $r < r_{EH}$ can likewise be set to zero by the same convention as in the flat Minkowski case for QCD. Regardless of their vacuum fluctuations, their vacuum averaged value $\varepsilon_{vac}$ must be zero, in spite of the uncertainty principle, else the Killing charges representing the background spacetime are not conserved.

Next we come to another important point in this picture. In modern cosmology, the notion of phase transitions plays a fundamental role. These involve SSB and contribute to the VED $\rho_V$. The consequence is that the effective vacuum potential $U(\phi)$ responsible for SSB has two phases [47] and takes on two vacuum states. In hadron physics, there is the bag constant $B \neq 0$ which represents an internal negative pressure $p = -B$ that subtends the hadron (Fig. 1, later). It is not the same as $<\rho_{vac}>$ for the background $\eta_{\mu\nu}$ in (7).

These multiple values of the gravitational APdS vacua are not space-time-dependent. Rather they are temperature-dependent. They occur because the vacuum exists at a finite temperature produced by curvature-induced quantum corrections in gauge theories with scalar fields [48]. Spacetime thus has a chemical potential and is temperature-dependent in these asymptotic metrics for temperature $T$. It is the presence of thermal matter (hadrons) that breaks the Lorentz invariance of these vacuum states [49].

Hence, during a phase transition in the early FLRW universe, the formation of hadrons has locally broken the Lorentz invariance of the global vacuum in (10). A local Lorentz boost and Poincaré translation from the outside of the hadron, into its interior, do not result in the same vacuum. This will become evident in Sect. 3.2.

Therefore, the fundamental basis for (9) and (10) cannot explain the existence of hadrons in the universe today. For this reason, we turn here to the original standard scalar-tensor theory [39-41] for an answer. The historical motivation for the JFBD theory was to create a time-dependent, variable gravitation constant $G=G(t)$. That is not the purpose here. Rather, the self-interacting scalar field $\phi$ will be regulated by the SSB process and must allow both $G$ and $\lambda$ to have two different states or values, one inside and one outside the hadron, that are temperature-dependent. EG in (9) cannot accomplish this.

### 3.2. Symmetry Breaking Potentials $U(\phi)$

There are many examples of symmetry breaking potentials $U(\phi)$. These include the well-known quartic Higgs potential for the Higgs complex doublet $\phi \to \Phi$

$$U(\Phi) = -\mu^2(\Phi^\dagger\Phi) + \varsigma(\Phi^\dagger\Phi)^2 \quad , \qquad (11)$$

where $\mu^2 > 0$ and $\zeta > 0$. (11) has minimum potential energy for $\Phi_{\min} = (0, v)^T/\sqrt{2}$ with $v = \sqrt{\mu^2/\varsigma}$. Viewed as a quantum field, $\Phi$ has the vacuum expectation value $<\Phi> = \Phi_{min}$. Following SSB, one finds $\Phi_{min} = (0, v+\eta(x))^T/\sqrt{2}$, indicating the appearance of the Higgs particle $\eta$. In order to determine the mass of $\eta$ one expands (11) about the minimum $\Phi_{min}$ and obtains

$$U(\eta) = U_o + \mu^2\eta^2 + \varsigma v\eta^3 + \frac{1}{4}\varsigma\eta^4 \quad , \qquad (12)$$

where $U_o = -\frac{1}{4}\mu^2 v^2$ is negative definite (wrong sign for solving the CCP), and $\eta$ acquires the mass $m_\eta = \sqrt{2\mu^2}$. Another example is the more general self-interacting quartic case

$$U(\phi) = U_o + \kappa\phi + \frac{1}{2}m^2\phi^2 + \varsigma v\phi^3 + \frac{c}{4!}\phi^4 \quad , \qquad (13)$$

investigated by [50,51] to examine the ground states of nonminimally coupled, fundamental quantized scalar fields $\phi$ in curved spacetime. $U_o$ is arbitrary. (13) is based upon the earlier work of T.D. Lee *et al.* [38,52-54] and Wilets [55] for modelling the quantum behavior of hadrons in bag theory

$$U^*(\sigma) = U_o + \frac{d}{2}T^*\sigma + \frac{a}{2}\sigma^2 + \frac{b}{3!}\sigma^3 + \frac{c}{4!}\sigma^4 \quad , \qquad (14)$$

where $\phi \to \sigma$ represents the self-interacting scalar $\sigma$-field as a nontopological soliton (NTS).[11] $U_o=B$ is the bag constant and is positive. The work of Friedberg, Lee, and Wilets (FLW) is reviewed in [55-58]. See also [59].

In all cases (12)-(14), $U_o$ represents a cosmological term,[12] and all are unrelated except that they represent the VED of the associated scalar field. The terms in $U(\phi)$ have a mass-dimension of four as required for renormalizability. In the case of (11)-(12), it is the addition of the Higgs scalar $\eta$ that makes the standard electroweak theory a renormalizable gauge theory. Also, the electroweak bosons obtain a mass as a result of their interaction with the Higgs field $\eta$ if it is present in the vacuum.

Note finally that (11)-(14) all have the same basic quartic form. The focus here will be on the hadron bag (14) in FLW theory. As pointed out by Creutz [60,61] the bag is an extended, composite object subject to non-local dynamics and not subject to perturbation theory. Following symmetry breaking, the soliton bag potential is depicted in Figure 1. The ground-state vacuum $<\rho_{vac}>$ at $<\sigma> = <\sigma_{vac}>$ is the APdS background defined by $\eta_{\mu\nu}$ in (7) as argued in Sect. 2 and is given by (1). The second vacuum state at $<\sigma> = 0$ is internal to the hadron and is given by the bag constant $B = \bar{\kappa}^{-1}\lambda_{Bag}$. $U_o = B$ in (14) is not a bare parameter determinable by a calculation, any more than $U_o$ in (12) derives a Higgs mass. $B$ is a fundamental parameter of FLW bag theory, is determinable by experimental hadron spectroscopy that models all hadrons, and is defined below.

---

[11] The asterisk in (14) is used to indicate that $d \neq 0$. $T^*\sigma$ is a chiral SB term that represents the cloud of pions surrounding the bag. $d = 0$ restores the symmetry.

[12] $U_o$ in (13)-(14) is an absolute number that is not experimentally measurable. This will become apparent in the definition of $B$ that follows.

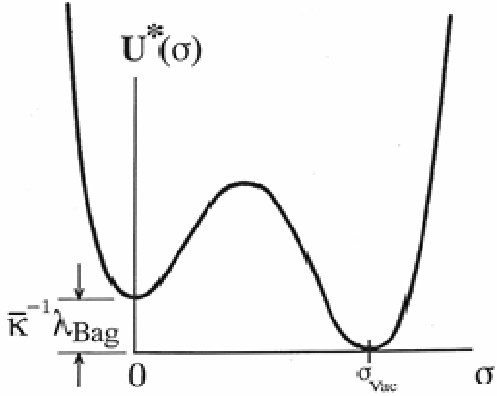

**Figure 1. The two vacuum states of the cosmological constant $\lambda=\lambda(\sigma)$ in the scalar-tensor model.** The scalar $\sigma$-field has undergone a phase transition and breaks the symmetry of the temperature-dependent vacuum, creating two vacuum states $\lambda_{Bag}$, and $\lambda_{vac}$. Inside the hadron at $<\sigma>=0$, $\lambda=\lambda_{Bag}$. Outside the hadron at $<\sigma>=<\sigma_{vac}>$, the gravitational ground-state energy density of the vacuum $E_{vac}$ is defined by the background metric $\eta_{\mu\nu}$ in (7) with $\lambda=\lambda_{F\text{-}L}$ for the Friedmann-Lemaître accelerating universe. Both are a de Sitter space.

On the other hand, quark and gluon confinement is generally attributed to the nonperturbative structure of the QCD vacuum. This is the basis for the MIT bag model [62,63] which first introduced $B$ and visualized hadrons as bubbles of perturbative (PT) vacuum immersed in the nonperturbative (NP) QCD vacuum. In that case, a truly NP VED in Yang-Mills theory has been derived [64,65]. The difference between PT and NP vacua is by definition $B$, provided of course that someday backreaction of the APdS spacetime is properly accounted for in QFT and QCD [66].

In what follows, we will show how to resolve CCP-1 in hadron physics using (14). This will be done in the fashion of a modified JFBD scalar $\sigma$ nonminimally coupled to the tensor field $g_{\mu\nu}$ in (9).

### 3.3. Merging Hadrons With Gravity

As discussed regarding (14), $\lambda$ in (9) is actually a potential term $\lambda(\sigma)$ that contributes to $U^*(\sigma)$. In that manner it couples to the SSB self-interacting quartic field $\sigma$ added to give QCD a scalar field [38] in the FLW bag model. However, here $\sigma$ will be treated as a gravitational field in order to address CCP-1.

Matter will be limited to hadron bags and $\lambda=\lambda(\sigma)$ will be moved to become a part of the bag potential in (14) as $U_o = B = \bar{\kappa}^{-1}\lambda_{Bag}$. This transposes $\lambda$ to the right-hand side[13] of (9) and gives the scalar-tensor field equations

$$R_{\mu\nu} - \frac{1}{2}g_{\mu\nu}R = -\kappa T^*_{\mu\nu} \quad , \tag{15}$$

$$T^*_{\mu\nu} = T^M_{\mu\nu} + T^\sigma_{\mu\nu} \quad , \tag{16}$$

$$\lambda_{Bag} = \bar{\kappa} B \quad , \tag{17}$$

where now $\lambda=\lambda(\sigma)$ contributes to the $\sigma$-field tensor $T^\sigma_{\mu\nu}$. The matter tensor is $T^M_{\mu\nu} = T_{\mu\nu}$ in (9), and their sum $T^*_{\mu\nu}$ in (16) is conserved by the Bianchi identities. We will derive $T^\sigma_{\mu\nu}$ below using scalar-tensor methods. Note that (17) resolves the mass dimensionality of $\lambda$ and $B$ in that both sides of the equation have mass dimension two.

This amounts to moving $\lambda$ about within the total Lagrangian $£ = T - U$ for the action involved, $S = S_{Gravity} + S_{Matter} + S_{G,M}$. Recalling that the Lagrangian for the FLW bag model $£_{FLW}$ is that for QCD ($£_q + £_C$) supplemented by the nonlinear $\sigma$-field $£^*_\sigma$ plus a quark-$\sigma$ mixing term $£_{q\text{-}\sigma}$,

$$£_{FLW} = £_q + £^*_\sigma + £_{q\text{-}\sigma} + £_C \quad , \tag{18}$$

the $£^*_\sigma$ term here will become the $\sigma$-field interaction term with scalar-tensor gravity $£^*_\sigma = £_{G,\sigma}$ in the total Lagrangian that includes a nonminimally coupled Einstein-Hilbert term $£_{\lambda JFBD}$ as

$$£_{Total} = £_{\lambda JFBD} + £^*_\sigma + £_q + £_{q\text{-}\sigma} + £_C \quad , \tag{19}$$

where

$$£^*_\sigma = \frac{1}{2}\nabla_\mu \sigma \nabla^\mu \sigma - U^*(\sigma) = £_{G,\sigma} \quad , \tag{20}$$

and $£_{\lambda JFBD}$ will be introduced shortly as (27).

The $\sigma$-field may be interpreted as a gluon condensate arising from nonlinear interactions of the color fields $£_C$ [56]. Regardless of its origin and composition, this scalar is the basis for the model under discussion.

For quarks $\psi$, scalar $\sigma$, and colored gluons $C$, these terms in (18) and (19) are

$$£_q = \bar{\psi}(i\gamma^\mu D_\mu - m)\psi \quad , \tag{21}$$

$$£_{q\text{-}\sigma} = -f\bar{\psi}\sigma\psi \quad , \tag{22}$$

$$£_C = -\frac{1}{4}\varepsilon(\sigma)\mathbf{F}_{\mu\nu}\mathbf{F}^{\mu\nu} - \frac{1}{2}g_s\bar{\psi}\lambda^c A^\mu_c \psi, \tag{23}$$

where counterterms are not shown. $m$ is the quark flavor mass matrix, $f$ the $\sigma$-quark coupling constant, $g_s$ the strong coupling, $\mathbf{F}_{\mu\nu}$ the non-Abelian gauge field tensor, $D_\mu$ the gauge-covariant derivative, and $\nabla_\mu$ the gravitation-covariant derivative (also in $\mathbf{F}_{\mu\nu}$) with the spin connection derivable upon solution of (15) above, defining the geodesics. $\varepsilon(\sigma)$ is the phenomenological dielectric function introduced by Lee *et al.* [52], where $\varepsilon(0)=1$ and $\varepsilon(\sigma_{vac})=0$ in order to guarantee color confinement. The $SU_3$ Gell-Mann matrices and structure factors are $\lambda_c$ and $f_{abc}$ respectively.

---

[13] Geometry in EG is determined by $g_{\mu\nu}$ – not which side of the equation $\lambda$ is on. Also $\lambda$ can only be introduced once, not both in (9) and (14) else there is double-counting.

Variation of (18) which neglects gravity in (19), using (20)-(23), gives the FLW equations of motion for $\sigma$ and $\psi$,

$$\Box\sigma = U^{*\prime}(\sigma) + f\,\bar{\psi}\psi \quad , \qquad (24)$$

$$(i\gamma^\mu D_\mu - m - f\sigma)\psi = 0 \quad , \qquad (25)$$

if one neglects the gluonic contribution (23). $\Box$ is the curved-space Laplace-Beltrami operator, and $U^{*\prime} = dU^*/d\sigma$ is

$$U^{*\prime} = \frac{d}{4}T^* + a\sigma + \frac{b}{2}\sigma^2 + \frac{c}{3!}\sigma^3 \quad . \qquad (26)$$

A variant adopts $d=0$ to simplify (26) when pion physics is not involved.

In the same fashion that $\lambda(\sigma)$ is a function of the $\sigma$-field, $\kappa$ is likewise as $\kappa(\sigma)$. For purposes here, the original JFBD ansatz $\kappa = \sigma^{-1}$ is adopted although there are others. This ansatz directly relates to (17). Taking into account (20), the nonminimally coupled scalar-tensor Lagrangian is

$$\mathcal{L}_{\lambda JFBD} = \frac{1}{2}\sqrt{-g}\left[-\sigma R + \frac{\Omega}{\sigma}\nabla_\mu\sigma\nabla^\mu\sigma - U^*(\sigma)\right] + 8\pi\mathcal{L}_{matter} \quad . \qquad (27)$$

The task now is to complete the scalar-tensor picture. The energy-momentum tensor in (16) is comprised of two terms. The first is the usual matter contribution $T^M_{\mu\nu}$ which includes all matter fields in the universe except gravitation,

$$T^M_{\mu\nu} = \frac{2}{\sqrt{-g}}\left[\frac{\partial(\sqrt{-g}L_M)}{\partial g^{\mu\nu}} - \partial^\alpha\frac{\partial(\sqrt{-g}L_M)}{\partial(\partial^\alpha g^{\mu\nu})}\right] \quad . \qquad (28)$$

It is thereby independent of the gravitational $\sigma$-field.[14]

The second term in (16) $T^\sigma_{\mu\nu} = \nabla_\mu\sigma\nabla_\nu\sigma - g_{\mu\nu}\mathcal{L}^*_\sigma$ is new and must include the effects of $\mathcal{L}_{G,\sigma}$ in (20). Consolidating all of the $\sigma$ terms and introducing a superscript "R" for renormalizable, we have in short-hand derivative notation

$$^R T^\sigma_{\mu\nu} = \sigma_{;\mu}\sigma_{;\nu} - \frac{1}{2}g_{\mu\nu}\sigma^{;\alpha}_{;\alpha} + g_{\mu\nu}U^*(\sigma) \quad . \qquad (29)$$

With (28) and (29), variation of (27) will now give the final equations of motion.

A principal assumption follows Brans and Dicke (BD). In order not to sacrifice the success of the principle of equivalence in Einstein's theory [10], only $g_{\mu\nu}$ and not $\sigma$ enters the equations of motion for matter consisting of particles and photons. The interchange of energy between matter and gravitation thus must follow geodesics as assumed by Einstein [67]. Therefore, the energy-momentum tensor for matter is assumed to be conserved in the standard fashion, $T^M_{\mu\nu;}{}^\mu = 0$.[15]

The derivation of $T^\sigma_{\mu\nu}$ is a textbook problem [67] except that the latter was a classical treatment following BD — both of which neglected $\lambda$, any potential $U^*(\sigma)$, and the renormalization restrictions on $U^*(\sigma)$ in (14).

The most general symmetric tensor of the form (29) which can be built up from terms each of which involves two derivatives of one or two scalar $\sigma$-fields, and $\sigma$ itself, is

$$T^\sigma_{\mu\nu} = A(\sigma)\sigma_{;\mu}\sigma_{;\nu} + B(\sigma)\delta_{\mu\nu}\sigma_{;\alpha}\sigma^{;\alpha} + C(\sigma)\sigma_{;\mu;\nu}$$
$$+ D(\sigma)\delta_{\mu\nu}\Box\sigma + E(\sigma)g_{\mu\nu}U^*(\sigma) \quad . \qquad (30)$$

We want to find the coefficients A, B, C, D, and E.

Taking the covariant divergence of (29) gives

$$^R T^\sigma{}^\mu_{\nu;\mu} = \sigma_{;\nu}\Box\sigma - \sigma^{;\mu}_{;\nu}\sigma_{;\mu} + U^{*\prime}(\sigma)\sigma_{;\nu} \quad , \qquad (31)$$

and that of (30) results in

$$T^\sigma{}^\mu_{\nu;\mu} = [A(\sigma) + B'(\sigma)]\sigma^{;\mu}_{;\nu}\sigma_{;\mu}$$
$$+ [A(\sigma) + D'(\sigma)]\sigma_{;\nu}\Box\sigma$$
$$+ [A(\sigma) + 2B(\sigma) + C'(\sigma)]\sigma^{;\mu}_{;\nu}\sigma_{;\mu}$$
$$+ [D(\sigma)](\Box\sigma)_{;\nu}$$
$$+ [C(\sigma)]\Box(\sigma_{;\nu})$$
$$+ [E(\sigma)U^{*\prime}(\sigma) + U^*(\sigma)E'(\sigma)]\sigma_{;\nu} \quad . \qquad (32)$$

Recalling the ansatz $\sigma \sim \kappa^{-1}$, next multiply (15) by $\sigma$ and take its divergence,

$$\left(R_{\mu\nu} - \frac{1}{2}g_{\mu\nu}R\right)^\mu_{;}\sigma + \left(R_{\mu\nu} - \frac{1}{2}g_{\mu\nu}R\right)\sigma^\mu_{;} = -8\pi T^{M\;\mu}_{\mu\nu;} - 8\pi T^{\sigma\;\mu}_{\mu\nu;} \quad . \qquad (33)$$

The first term on the l.h.s. of (33) is zero by the Bianchi identities; the first on the r.h.s is zero by the principle of equivalence. The net result is

$$\left(R_{\mu\nu} - \frac{1}{2}g_{\mu\nu}R\right)\sigma^\mu_{;} = -8\pi T^{\sigma\;\mu}_{\mu\nu;} \quad . \qquad (34)$$

Using an identity [67] involving the Riemann tensor $R^\gamma_{\alpha\nu\beta}$, the first term in (34) is

$$R_{\mu\nu}\sigma^{;\mu} = \sigma^{;\alpha}_{;\alpha;\nu} - \sigma^{;\alpha}_{;\nu;\alpha} = (\Box\sigma)_{;\nu} - \Box(\sigma_{;\nu}) \quad . \qquad (35)$$

Take the trace of (15) and (16) for $R$. Next modify (24) to include the gravitational coupling with $\sigma$ (still assuming $f = 0$) to produce the trace for $T^M$. Lastly obtain the remaining trace for $T^\sigma$ from (30). These three traces are

$$R = \kappa T^M + \kappa T^\sigma \quad , \qquad (36)$$

$$T^M = 2\kappa_1^{-1}[\Box\sigma + U^{*\prime}(\sigma)] \quad , \qquad (37)$$

$$T^\sigma = [A(\sigma) + 4B(\sigma)]\sigma^{;\alpha}\sigma_{;\alpha} + [C(\sigma) + 4D(\sigma)]\Box\sigma + 4[E(\sigma)U^*(\sigma)]. \quad (38)$$

---

[14] In (28), $L_M \to \mathcal{L}_{matter}$.

[15] Exceptions can be made, but will not be entertained here. See the earlier study in Ref. 68.

It follows that the collective trace for $R$ is

$$R = 2\kappa\kappa_1^{-1}[\Box\sigma + U^{*\prime}(\sigma)] + \kappa\{[A(\sigma)+4B(\sigma)]\sigma^{;\alpha}\sigma_{;\alpha} \\ + [C(\sigma)+4D(\sigma)]\Box\sigma + 4[E(\sigma)U^{*}(\sigma)]\} \quad . \quad (39)$$

Placing (39) into the left-hand-side of (34) with some re-arrangement gives

$$\left(R_{\mu\nu} - \frac{1}{2}g_{\mu\nu}R\right)\sigma_{;}^{\mu} = -\tfrac{1}{2}\kappa[A'(\sigma)+4B(\sigma)]\sigma_{;}^{\mu}\sigma_{;\nu}\sigma_{;\mu}$$
$$-\tfrac{1}{2}\kappa[2\kappa_1^{-1}+C(\sigma)+D(\sigma)]\sigma_{;\nu}\Box\sigma$$
$$+ [0]\sigma_{;}^{\mu}{}_{;\nu}\sigma_{;\mu}$$
$$+ [1](\Box\sigma)_{;\nu}$$
$$+ [-1]\Box(\sigma_{;\nu})$$
$$-\tfrac{1}{2}\kappa[2\kappa_1^{-1}U^{*\prime}(\sigma)+4E(\sigma)U^{*}(\sigma)]\sigma_{;\nu} \quad . \quad (40)$$

In order that (34) be true, the bracketted coefficients in (32) and (40) must be equal term by term. Renormalization problems created by $E(\sigma)$ are addressed in Ref. 68. These include the insolvability of a quintic and the Galois-Abel theorem.[16] Finally, one encounters the result $A(\sigma) = \tfrac{1}{2}\kappa[\kappa_1^{-1} - 3/2]$ which prompts the definition

$$\Omega = \kappa_1^{-1} - \frac{3}{2} \quad , \quad (41)$$

whereby $\kappa_1$ in (37) is

$$\kappa_1 = \frac{2}{3+2\Omega} \quad . \quad (42)$$

The desired energy-momentum tensor for the $\sigma$-field follows as

$$\kappa T^{\sigma}_{\mu\nu} = \frac{\Omega}{\sigma^2}[\sigma_{;\mu}\sigma_{;\nu} - \frac{1}{2}g_{\mu\nu}\sigma_{;\alpha}\sigma^{;\alpha}] - \frac{1}{\sigma}[\sigma_{;\mu}\sigma_{;\nu} - g_{\mu\nu}\Box\sigma]$$
$$- \frac{1}{\sigma}[g_{\mu\nu}U^{*}(\sigma)] \quad . \quad (43)$$

Inserting (43) into (15) and (16) gives the full field equations

$$(R_{\mu\nu} - \frac{1}{2}g_{\mu\nu}R) = -\frac{8\pi}{\sigma}T^{M}_{\mu\nu} - \frac{\Omega}{\sigma^2}[\sigma_{;\mu}\sigma_{;\nu} - \frac{1}{2}g_{\mu\nu}\sigma_{;\alpha}\sigma^{;\alpha}]$$
$$- \frac{1}{\sigma}[\sigma_{;\mu}\sigma_{;\nu} - g_{\mu\nu}\Box\sigma] - \frac{1}{\sigma}[g_{\mu\nu}U^{*}(\sigma)] \quad , \quad (44)$$

while (43) in (37) gives the scalar wave equation (for $f=0$) for the $\sigma$-field

$$\Box\sigma = \frac{8\pi}{3+2\Omega}T^{*} + U^{*\prime}(\sigma) \quad , \quad (45)$$

where $\Omega = (\kappa_1^{-1} - 3/2)$ and $\kappa_1$ is the source of $\sigma$-coupling to the traditional trace $T^M$ in JFBD theory. There is now coupling to the trace $T^*$ in (45) compared to (24). If $\Omega = -3/2$, (44) is a conformally mapped set of Einstein field equations.

### 3.4. Characterizing Scalar-Tensor Gravity and Hadrons

In order to visualize the results (44), (45), and (25), Figure 2 illustrates this scalar-tensor approach to CCP-1.

---
[16] Hence $E(\sigma)$ must be $E(\sigma) = 1$.

*Bag Boundary Conditions.* Bag surface boundary conditions are discussed in Ref. 69, p. 103, where the following can be adopted: $\mathbf{F}_{\mu\nu}n^{\nu} = 0$ for gauge fields; and $\psi = 0$ for quark fields. As a problem in bubble dynamics, one uses $-\tfrac{1}{4}\mathbf{F}^{\mu\nu}\mathbf{F}_{\mu\nu} = \Sigma J^{-1}\nabla_{\nu}(Jn^{\nu}) + B$ for a quark current $J$ and surface tension $\Sigma$. Alternatively the more recent Lunev-Pavlovsky bag with a singular Yang-Mills solution on the bag surface [70-73] can be utilized (also probably eliminating the need for $\varepsilon(\sigma)$ in (23), a point that is yet to be addressed).

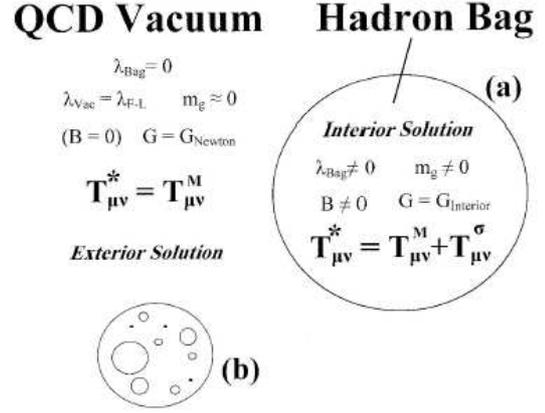

**Figure 2. The existence of two vacuum states for $\lambda=\lambda(\sigma)$ characterized by equations (16), (44)-(45), and (47).** The exterior is traditional Einstein gravity where $\lambda=\lambda_{F-L}$. (a) A single hadron bag is depicted with $B = \bar{\kappa}^{-1}\lambda_{Bag}$. (b) The general interior solution is depicted as a many-bag problem using a Swiss-cheese (modified Einstein-Straus) model with zero pressure on the bag surfaces. Applicable boundary conditions are in [69-73].

From the dimensionality of $U^{*}(\sigma)$ in (14), we see that $a$ has mass-dimension two or $m^2$. Taking the derivative $U^{*\prime}(\sigma)$ as in (26) along with (45), the $\sigma$-field has mass

$$m_{\sigma} = \sqrt{a} \quad . \quad (46)$$

Therefore it is a short-range field with only short-range interaction. (45) can be re-written

$$(\Box - m_{\sigma}^2)\sigma = \delta U^{*\prime}(\sigma) + \frac{8\pi}{3+2\Omega}T^M + f\bar{\psi}\psi \quad , \quad (47)$$

where $\delta U^{*\prime}$ is the remainder of (26) after moving the $a\sigma$ term to the left-hand side. Hence a static solution must have a Yukawa cutoff $\sigma \sim (e^{-\mu r})$ where $\mu \sim m_{\sigma}$.

This is characterized in Figure 2 by indicating that the energy-momentum tensor $T^{\sigma}_{\mu\nu}$ is confined to the hadron which is consistent with the original conjecture of FLW that the $\sigma$-field be related to confinement.

*Bag Interior Conditions.* By virtue of the bag condition (17), several new features come into play. First, this relation is specific to the interior of the bag. Second, the BD ansatz $\sigma \sim \kappa^{-1} \sim G^{-1}$ is now tied to the cosmological parameter $\lambda$ too.[17] This means that when the phase transition and SSB occur, there exist two de Sitter spaces in Figures 1 and 2. Both $\lambda$ and $G$ can differ between the two vacuum states. When $B=0$ (no bags), $\lambda \to \lambda_{vac}$ in Figure 1, restoring itself to the ground-state vacuum of the APdS background (Sect. 2). Similarly, $G \sim \sigma^{-1}$ is not guaranteed in theory to be the same in the hadron interior as $G_{Newton}$ outside. This is a matter for experimental investigation (Sect. 4.2 below).

*Summary.* Semi-classically speaking, the scalar-tensor theory of gravity in the presence of the QCD Lagrangian representing FLW bag theory in (18) and (19) has no apparent problem associated with the existence of two vacuum states, one in the exterior and one in the interior of hadrons.

This model has been done entirely in de Sitter space, whereby the principle of compatible asymptotic Killing charges (Sect. 2) has not been broken. Thus, CCP-1 does not appear to apply to scalar-tensor gravity when non-minimally coupled to hadron physics. The model allows for two vacuum states, as bags indicate experimentally. One is the F-L ground state ($\lambda=\lambda_{F-L}$) and the other is the hadron interior ($\lambda=\lambda_{Bag}$).

However, the issue of vacuum stability for this model is a crucial assumption because one can argue that it is unstable to radiative corrections. But radiative corrections have long been suggested as the origin of SSB to begin with [74]. These similarly are important for dynamical SB (DSB) models as well [75,76]. Since SSB has been adopted for the basic quartic potential $U^*(\sigma)$ in (13)-(14), then there has been an implicit assumption that the scalar-tensor configuration presented here is stable to radiative corrections. Vacuum stability of this model is a subject for further study, in particular when QCD confinement is more thoroughly understood.

## 4. Cosmological Event Horizons, Finite Temperature, and Experiment

The connection between $\lambda$ and graviton mass (Appendix) has a bearing upon identifying the APdS spacetime as the ground state vacuum for the CCP with its associated Killing charges in the AD formalism, in order to rectify CCP-2 in Sect. 2. Because there exists the G-H event horizon in such cosmological spacetimes, any association of $\lambda$ with a graviton mass is very pertinent even if it is gauge dependent.

In the Appendix as (67), it is shown in the weak-field approximation that a graviton mass $m_g = \sqrt{\lambda/3}$ is associated with a small $\lambda$ such as (1) for de Sitter spacetime. It is equivalent to the surface gravity $\kappa_C = \sqrt{\lambda/3}$ found by G-H [24]. These in turn relate directly to the radius of the event horizon $r_{EH}$ defined by the singularity in (2) and (3) at $r_{EH} = \kappa_C^{-1}$.

Next, finite temperature effects must be discussed and this is done below in Sect. 4.1. Experimental aspects follow in Sect. 4.2 and are recapitulated in Table 1.

### 4.1. Finite Temperature Effects

A digression on the effect of finite temperature $T$ upon $U^*(\sigma)$ is pertinent because it is relevant to experiment. The subject is also pertinent to the basic concept of a temperature-dependent spacetime in gravitation theory, and equally so to the topic of cosmological event horizons.

The subject is treated in the usual fashion [77-79]. The classical, zero temperature potential $U^*(\sigma)$ in (14) becomes $V^*(\sigma) = U^*(\sigma) + V_S(\sigma,T) + V_F(\sigma,T,\mu)$. This involves scalar $V_S$ and fermionic $V_F$ correction terms for chemical potential $\mu$, by shifting $\sigma$ as $\sigma = \sigma' + v(T)$. The result is a temperature-dependent cosmological bag parameter [80] $\lambda_{Bag} = \lambda_{Bag}(\mu,T) = \bar{\kappa} B(\mu,T)$ which decreases with increasing temperature $T$ until the bag in Figure 2 dissolves and symmetry is restored ($B=0$) in Figure 1.

In such a case and in simplest form [81], the bag model equations of state are

$$\varepsilon(T) = k_{SB}T^4 + B \quad , \quad (48)$$

$$p(T) = \frac{1}{3}k_{SB}T^4 - B \quad , \quad (49)$$

$$k_{SB} = \frac{\pi^2}{30}(d_B + \frac{7}{8}d_F) \quad , \quad (50)$$

where energy density $\varepsilon$ and pressure $p$ now have a temperature dependence ($T \neq 0$). The Stefan-Boltzmann (SB) constant $k_{SB}$ is a function of the degeneracy factors $d_B$ for bosons (gluons) and $d_F$ for fermions (quarks and antiquarks). The absence of the baryonic chemical potential $\mu$ in (48) is a valid approximation for ongoing experiments involving nucleus-nucleus collisions. All are relevant to quark-hadron phase transitions and the quark-gluon plasma (QGP).

### 4.2. Experimental Aspects

As mentioned previously (Sect. 3), EG appears to be the correct theory of gravity above 1 mm. The subject here is below that scale in the Large Hadron Collider (LHC) realm of particle and nuclear physics. Granted, the treatment in this paper has neglected the standard model in order to present a tractable discussion of hadrons, gravity, and the CCPs.

*Within the hadron bag.* Here one has $m_g \neq 0$ due to (17) and (67). Adopting a simplified view of the hadron interior and a bag constant value from one of the conventional bag models, the MIT bag [62-63] where $B^{1/4} = 146 MeV$ or $B = 60 MeV\, fm^{-3}$, then $\lambda_{Bag} = \bar{\kappa} B = 2 \times 10^{-13} cm^{-2}$ follows from (17).[18] Using (67) in the Appendix, a graviton mass $m_g = 2.6 \times 10^{-7} cm^{-1}$ or $5.2 \times 10^{-12} eV$ is found within the bag. Although this appears to represent a Compton wavelength of $m_g^{-1} \sim 4 \times 10^6 cm$ or range of ½ $m_g^{-1} \sim 2 \times 10^6 cm$, it is derived from $\lambda_{Bag}$ and is only applicable for the interior solution. This is depicted in Figure 2. It has no range outside of the bag where $\lambda_{Bag} = 0$.

---

[17] $\lambda$ as a VED is thereby possibly related to the origin of $G$.

A similar calculation for the Yang-Mills condensate [64,65] $B_{YM} \sim 0.02 GeV^4$ gives $\lambda_{YM} \sim 8.7 \times 10^{-12} cm^{-2}$ and $m_g \sim 1.7 \times 10^{-6} cm^{-1}$ or $3 \times 10^{-11} eV$, and ½ $m_g^{-1} \sim 3.5 \times 10^5 cm$.

Regarding $G$, adopting $G_{Bag} = G_{Newton}$ is the sensible assumption to make. However, $G_{Bag}$ is a free parameter, independent of $B$. It has never been experimentally measured. For any $B$ determined in Table 1, $G_{Bag}$ can be anything except zero.[19]

*The bag per se.* The $\sigma$-field has a mass (46) in Table 1 subject to experimental measurement, perhaps at the LHC in scalar gluon jets related to ongoing boson searches to complete the Standard Model [82,83]. It is conceivable that evidence for both the Higgs boson and the scalar $\sigma$-field used here for the bag can be found.[20]

*External to the hadron.* By taking the well-known JFBD limit $\Omega \to \infty$ in (44) and (45), we in fact obtain Einstein gravity (for exceptions see Ref. 45) due to the experimental limits [42-44]. The small graviton mass $m_g$ in (67), on the other hand, results in a finite-range gravity whose mass is $m_g \sim 0.6 \times 10^{-28} cm^{-1}$ or $1.1 \times 10^{-33} eV$. This follows from the vacuum energy density $(\sim 2 \times 10^{-3} eV)^4$ which is equivalent to $\lambda \sim 10^{-56} cm^{-2}$, for the de Sitter background $\eta_{\mu\nu}$ in (7) for the F-L accelerating universe [6,7].

Obviously, $G = G_{Newton}$ in the exterior.

*Summary.* The results for this model are as follows. In the exterior we have a graviton mass $m_g \sim 0.6 \times 10^{-28} cm^{-1}$ and a range of ½ $m_g^{-1} \sim 8 \times 10^{27} cm$ which is approximately the Hubble radius. That is, gravitation outside of the bag is finite-ranged reaching to the G-H cosmological event horizon $r_{EH} = \kappa_C^{-1}$. Gravitation within the bag is short-ranged.

Clearly the sign of $\lambda$ must be positive (de Sitter space) in (67) in order that an imaginary mass not be possible.

The latter represents an unstable condition with pathological problems such as tachyons and negative probability. (67) is a physical argument against such a circumstance.

**Table 1. Summary of the masses, VED's, and $\lambda$'s in spacetime.**

| Spacetime Region | $m_g$ ($cm^{-1}$) | $m_g$ ($eV$) | $m_\sigma$ ($GeV$) | VED, $B$ ($GeV$)$^4$ | $\lambda$ ($cm^{-2}$) |
|---|---|---|---|---|---|
| *Hadron exterior* $\lambda \equiv \lambda_{F-L} \neq 0$ | $0.6 \times 10^{-28}$ | $1.1 \times 10^{-33}$ | | $2 \times 10^{-47}$ | $0.7 \times 10^{-56}$ |
| *Hadron interior* | | | | | |
| MIT bag [62] | $2.6 \times 10^{-7}$ | $5.2 \times 10^{-12}$ | $\sqrt{a}$ | 0.0045 | $2 \times 10^{-13}$ |
| Y-M cluster[64] | $1.7 \times 10^{-6}$ | $3 \times 10^{-11}$ | $\sqrt{a}$ | 0.02 | $9 \times 10^{-12}$ |

## 5. Assumptions and Postulates

At this point the fundamental postulates that have been made are summarized. These have been discussed and alluded to throughout but are now recapitulated.

1) Einstein gravity is the true theory of gravity at length scales above 1 mm.

2) The gravitational field $g_{\mu\nu}$ couples minimally and universally to all of the fields of the Standard Model, as does Einstein gravity [10]. However, $g_{\mu\nu}$ also couples nonminimally to the composite features of $\pounds_{FLW}$ and $\pounds_{QCD}$. The $\pounds_{FLW}$ term represents hadron physics which includes QCD in the exact limit $\pounds_{FLW} \to \pounds_{QCD}$ (see Ref. 56, p. 19). The JFBD ansatz $\kappa = \sigma^{-1}$ is assumed.

3) The nonlinear self-interacting scalar $\sigma$-field represented by Lagrangian $\pounds^*_\sigma$ is a gravitational field, because it couples universally to all hadronic matter. Since $\sigma$ has a mass (46) it and $T^\sigma_{\mu\nu}$ in (16) have a cutoff and are confined to the hadron in Fig. 2a.

4) General covariance is necessary in order that the Bianchi identities determine conservation of energy-momentum from $T^*_{\mu\nu}$ in (15). However, in the hadron exterior, $T^*_{\mu\nu} \to T^M_{\mu\nu}$. That means matter follows Einstein geodesics and obeys the principle of equivalence as expected there.

5) Stability must be assumed for $\delta T_{\mu\nu}$ in the Appendix. Use of the harmonic gauge, $f_\mu = 0$ in (53), suppresses the vector gravitons and manifests a tiny graviton mass $m_g = \sqrt{2\lambda/3}$, but breaks general covariance. The consequence is not measurable within the observer's cosmological event horizon.

---

[18] In terms of units, the following conversions are helpful: 1 $MeV^4$ = $2.3201 \times 10^5$ g $cm^{-3}$, then $\bar{\kappa} = 1.8658 \times 10^{-27}$ cm $g^{-1}$ or $\bar{\kappa} = 4.3288 \times 10^{-22}$ $cm^{-2}\, MeV^{-4}$. Thus $\lambda_{Bag} = \bar{\kappa} B = 2 \times 10^{-13}$ $cm^{-2}$ for $B[g\, cm^{-3}] = 2.3201 \times 10^5$ $B[MeV^4]$. This assumes $G = G_{Newton}$.

[19] This would move the Planck mass, $M_{Planck} \sim G_{Bag}^{-1/2}$.

[20] The recent suggestion of Friedberg and Lee [84] that the Higgs itself is composite and not elementary is very relevant to the experimental interpretations. See their references to earlier work on this subject [e.g. 85] and Lee's website.

6) Temperature-dependent quantum vacuum fluctuations result in a broken vacuum symmetry, producing two distinct vacua containing two different vacuum energy densities $\lambda_{F-L}$ and $\lambda_{Bag}$. Lorentz and Poincaré invariance are broken by $T^{\sigma}_{\mu\nu}$ in the interior of hadrons. Because $\lambda=\lambda(T)$, this broken symmetry is subject to restoration.

7) The stability of the bag is assured by the vacuum energy density $B$ which is a negative vacuum pressure. Similarly, the scalar-tensor representation of the hadron interior is stable against radiative corrections.

8) The principle of compatible asymptotic states (Killing charges) is assumed. This means that the global energies of flat ADM metrics are not compatible with those of APdS metrics. ADM energies cannot be consistently compared globally with AD energies in the definition of ground-state vacua for de Sitter space, lest infinities be introduced. Hence, derivations in flat Minkowski space are not relevant to the CCP if (1) is accepted as evidence for $\lambda$ contributing to the acceleration of the universe in FLRW cosmology whose current phase is an APdS metric.

## 6. Conclusions

A tenable model for the origin of hadron bound states in bag theory has been shown to derive from the cosmological constant $\lambda$ in scalar-tensor gravity, noting that the familiar Higgs mechanism does not account for the mass of composite particles such as hadrons. The bag model of Friedberg, Lee, and Wilets (FLW) is used instead.

According to the development in Section 3, a scalar-tensor treatment of $g_{\mu\nu}$ nonminimally coupled to hadrons using the nonlinear self-interacting scalar field $\sigma$ results in a model of gravity that has two different ground-state vacua. Such a theory exists and resolves CCP-1 adopting the assumptions made here. Experiment and theory must eventually settle the differences between the MIT bag model and the Yang-Mills condensate solution for bag constant $B$ in Sect. 3.2, but this does not alter these results. As a model, this point of view represents a tenable strategy for reconsidering CCP-1 and CCP-2 from hadron physics to cosmology. Without directly relating the bag constant to the global energy in APdS spacetime, any of the other proposed "solutions" of the CCP(s) are incomplete.

There is little surprise regarding CCP-2 for the large disparity between ground-state VEDs when derivations in flat Minkowski spacetime are being directly compared with those from APdS in cosmological gravity. This breaks the principle of compatible asymptotic states, by comparing energies derived from spacetimes that have entirely different Killing charges and global energy properties. A great deal of work on APdS structure and its relation to VEDs is therefore required before we will truly understand the CCP.

Finally, conventional massive gravity $m_g = m_{PF}$ has not been used in the strategy proposed here to address the CCP ($m_{PF} = 0$). This investigation involves only $\lambda$ and its relationship to asymptotic infinity, with a graviton mass $m_g$ (65) and (67) that manifests itself by suppressing the vector gravitons $f_\mu$ in (53). In this study, $m_g$ arises instead by introducing $\lambda \neq 0$ into the well-known Regge-Wheeler problem (see Appendix).

## 7. Acknowledgements

The author would like to acknowledge helpful communications with B. Tekin, as well as O.V. Pavlovsky, and S.J. Aldersley.

## Appendix: The Cosmological Constant as a Gravitational Mass

It was shown some time ago by this author [86] that the cosmological term $\lambda$ in General Relativity can be interpreted as a graviton mass in the weak-field approximation, by introducing $\lambda \neq 0$ into the Regge-Wheeler problem.

Note with caution that an unqualified graviton mass is beset with numerous problems in QFT. $T_{\mu\nu}$ in (9) has an admixture of Spin 2, three Spin-1, and two Spin-0 components. These can lead to loss of unitarity, negative energy states, and ghosts[21]. They possess 16 degrees of freedom which are reduced to 10 by energy conservation, then to six (6) by symmetry ($T_{\mu\nu} = T_{\nu\mu}$) leaving Spin-2, Spin-1, and two Spin-0 helicities. Invoking a Lorentz-type gauge condition such as $f_\mu=0$ in (53) that follows in Sect. A.2, eliminates the Spin-1 vector gravitons. A pure Spin-2 plus a Spin-0 that is coupled to the trace $T = T_\mu^\mu$ remains. If trace-free (such as empty space), pure Spin-2 remains. If the trace and the Spin-0 term remain, a scalar-tensor theory survives [89] which is what we assume here to begin with.

The above Spin-2 problems in fact motivated Pauli and Fierz [90] to introduce a graviton mass $m_{PF}$ (with $\lambda=0$) patterned after the Klein-Gordon mass in particle physics using the Lagrangian $\pounds_{PF} = \frac{1}{4} m_{PF}^2 (h_{\mu\nu} h^{\mu\nu} - h^\mu_\mu{}^2)$ so as to achieve a massive graviton without loss of unitarity. $\pounds_{PF}$ and its associated mass $m_{PF}$ are never introduced in this study ($m_{PF} \equiv 0$). There also has been controversy involving $m_{PF}$ graviton propagators in the massless limit $m_{PF} \to 0$ known as the vDVZ discontinuity. This has been resolved only recently (Sect. A.4 below).

In Sect. A.1 and A.2 the graviton mass associated with $\lambda$ is derived. Sect. A.3 shows that there is no hidden $\lambda$–term in the curved-space Laplace-Beltrami term that cancels out the result. Sect. A.4 discusses $m_{PF}$, the vDVZ discontinuity, its resolution, and unitarity of $\lambda$ as a graviton mass. Lastly, the issue of relaxing the assumption $f_\mu=0$ in (53) that suppresses the vector gravitons, and why is also discussed.

### A.1 Weak-Field Limit, Schwarzschild-de-Sitter Metric

The curved background first adopted is the Schwarzschild-de-Sitter (SdS) metric (2)-(3) applied to the Regge-Wheeler-Zerilli (RWZ) problem [91-94] for gravitational radiation perturbations produced by a particle falling onto a large mass $M^*$ with $\lambda=0$.

One considers a small perturbative expansion of EG (9) about the known exact solution $\eta_{\mu\nu}$ given in (2)-(3) subject to the boundary condition that $g_{\mu\nu}$ becomes $\eta_{\mu\nu}$ as $r \to \infty$. The metric tensor $g_{\mu\nu}$ is thus assumed to be (7) $g_{\mu\nu} = \eta_{\mu\nu} + h_{\mu\nu}$ where $h_{\mu\nu}$ is the dynamic perturbation such that $h_{\mu\nu} \ll \eta_{\mu\nu} = g_{\mu\nu}^{(0)}$.

The wave equation for gravitational radiation $h_{\mu\nu}$ follows as (56) below, derived exactly from the RWZ formalism. Perturbation analysis of (9) for a stable background $\eta_{\mu\nu} = g_{\mu\nu}^{(0)}$ produces the following

$$[h_{\mu\nu;\alpha}{}^{;\alpha} - h_{\mu\alpha;\nu}{}^{;\alpha} - h_{\nu\alpha;\mu}{}^{;\alpha} + h_\alpha{}^\alpha{}_{;\mu;\nu}] + \eta_{\mu\nu}[h_{\alpha\gamma}{}^{;\alpha;\gamma} - h_\alpha{}^\alpha{}_{;\gamma}{}^{;\gamma}]$$
$$+ h_{\mu\nu}(R - 2\lambda) - \eta_{\mu\nu} h_{\alpha\beta} R^{\alpha\beta} = -2\kappa \delta T_{\mu\nu} \quad . \quad (51)$$

Stability must be assumed in order that $\delta T_{\mu\nu}$ is small. This equation can be simplified by defining the function (introduced by Einstein himself)

$$\bar{h}_{\mu\nu} \equiv h_{\mu\nu} - \frac{1}{2}\eta_{\mu\nu} h \quad (52)$$

and its divergence

$$f_\mu \equiv \bar{h}_{\mu\nu}{}^{;\nu} \quad . \quad (53)$$

Substituting (52) and (53) into (51) and re-grouping terms gives

$$\bar{h}_{\mu\nu;\alpha}{}^{;\alpha} - (f_{\mu;\nu} + f_{\nu;\mu}) + \eta_{\mu\nu} f_\alpha{}^{;\alpha} - 2\bar{h}_{\alpha\beta} R^\alpha{}_{\mu\nu}{}^\beta - \bar{h}_{\mu\alpha} R^\alpha{}_\nu - \bar{h}_{\nu\alpha} R^\alpha{}_\mu$$
$$+ h_{\mu\nu}(R - 2\lambda) - \eta_{\mu\nu} h_{\alpha\beta} R^{\alpha\beta} = -2\kappa \delta T_{\mu\nu} \quad . \quad (54)$$

Now impose the Hilbert-Einstein-de-Donder gauge which sets (53) to zero ($f_\mu = 0$), and suppresses the vector gravitons. Wave equation (54) reduces to

$$\bar{h}_{\mu\nu;\alpha}{}^{;\alpha} - 2\bar{h}_{\alpha\beta} R^\alpha{}_{\mu\nu}{}^\beta - \bar{h}_{\mu\alpha} R^\alpha{}_\nu - \bar{h}_{\nu\alpha} R^\alpha{}_\mu$$
$$- \eta_{\mu\nu} h_{\alpha\beta} R^{\alpha\beta} + h_{\mu\nu}(R - 2\lambda) = -2\kappa \delta T_{\mu\nu} \quad . \quad (55)$$

In an empty ($T_{\mu\nu} = 0$), Ricci-flat ($R_{\mu\nu} = 0$) space without $\lambda$ ($R = 4\lambda = 0$), (55) further reduces to

$$\bar{h}_{\mu\nu;\alpha}{}^{;\alpha} - 2 R^\alpha{}_{\mu\nu}{}^\beta \bar{h}_{\alpha\beta} = -2\kappa \delta T_{\mu\nu} \quad , \quad (56)$$

which is the starting point for the RWZ formalism.

### A.2 Weak-Field Limit, de Sitter Metric

The Schwarzschild character of the RWZ problem above will now be relaxed, with $\eta_{\mu\nu}$ again diagonal, but $m=0$ and $\lambda \neq 0$ in (2) and (3). The wave equation of paramount importance will follow as (64).

We know that the trace of the field equations (9) gives $4\lambda - R = -\kappa T$, whereby they become

$$R_{\mu\nu} - \lambda g_{\mu\nu} = -\kappa[T_{\mu\nu} - \frac{1}{2} g_{\mu\nu} T] \quad . \quad (57)$$

For an empty space ($T_{\mu\nu} = 0$ and $T = 0$), (57) reduces to de Sitter space

$$R_{\mu\nu} = \lambda g_{\mu\nu} \quad . \quad (58)$$

and the trace to $R = 4\lambda$.

---

[21] Since a ghost has a negative degree of freedom (DOF), more ghosts must be introduced due to perturbative Feynman rules that over-count the correct degrees of freedom [87, 88].

Substitution of $R$ and $R_{\mu\nu}$ from (58) into (55) using (52) shows that the contributions due to $\lambda \neq 0$ are of second order in $h_{\mu\nu}$. Neglecting these terms (particularly if $\lambda$ is very, very small) simplifies (55) to

$$\bar{h}_{\mu\nu;\alpha}{}^{;\alpha} - 2R^{\alpha}{}_{\mu\nu}{}^{\beta}\bar{h}_{\alpha\beta} = -2\kappa\delta T_{\mu\nu} \quad . \quad (59)$$

One can arrive at (59) to first order in $h_{\mu\nu}$ by using $g_{\mu\nu}$ as a raising and lowering operator rather than the background $\eta_{\mu\nu}$ – a result which incorrectly leads some to the conclusion that $\lambda$ terms cancel in the gravitational wave equation.

Note with caution that (59) and the RWZ equation (56) are not the same wave equation. Overtly, the cosmological terms have vanished from (59), just like (56) where $\lambda$ was assumed in the RWZ problem to be nonexistent in the first place. However, the character of the Riemann tensor $R^{\alpha}{}_{\mu\nu}{}^{\beta}$ is significantly different in these two relations.

Simplifying the SdS metric by setting the central mass $M^*$ in $\eta_{\mu\nu}$ to zero, produces the de Sitter space (58) of constant curvature $K = 1/R^2$, where we can focus on the effect of $\lambda$. The Riemann tensor is now

$$R_{\mu\nu\delta} = +K(g_{\gamma\nu}g_{\mu\delta} - g_{\gamma\delta}g_{\mu\nu}) \quad , \quad (60)$$

and reverts to

$$R^{\alpha}{}_{\mu\nu}{}^{\beta} = +K(g^{\alpha}{}_{\nu}g_{\mu}{}^{\beta} - g^{\alpha\beta}g_{\mu\nu}) \quad , \quad (61)$$

for use in (59). This substitution (raising and lowering with $\eta_{\mu\nu}$) into (59) next gives $K$ and $\lambda$ term contributions

$$-2K[(\bar{h}_{\mu\nu} - \eta_{\mu\nu}\bar{h}) + (\bar{h}_{\alpha\mu}h^{\alpha}{}_{\nu} + \bar{h}_{\nu\beta}h^{\beta}{}_{\mu} - \bar{h}h_{\mu\nu} - \eta_{\mu\nu}h^{\alpha\beta}\bar{h}_{\alpha\beta})]$$
$$+ [2h_{\mu\alpha}\bar{h}^{\alpha}{}_{\nu} + \eta_{\mu\nu}h^2_{\alpha\beta}] \quad , \quad (62)$$

to second order in $h_{\mu\nu}$. Recalling that curvature $K$ is related to $\lambda$ by $K = \lambda/3$, substitution of (62) back into (59) gives to first order

$$\bar{h}_{\mu\nu;\alpha}{}^{;\alpha} - \frac{2}{3}\lambda\bar{h}_{\mu\nu} + \frac{2}{3}\lambda\eta_{\mu\nu}\bar{h} = -2\kappa\delta T_{\mu\nu} \quad . \quad (63)$$

There is no cancellation of the $\lambda$ contributions to first order. Noting from (52) that $\bar{h} = h(1-\frac{1}{2}\eta)$, then a traceless gauge $\bar{h} = 0$ means either that $h = 0$ or $\eta = 2$. Since $\eta = 4$, (63) reduces to

$$\bar{h}_{\mu\nu;\alpha}{}^{;\alpha} - \frac{2}{3}\lambda\bar{h}_{\mu\nu} = -2\kappa\delta T_{\mu\nu} \quad (64)$$

in a traceless Hilbert-Einstein-de Donder gauge where $\bar{h}_{\mu\nu}{}^{;\nu} = 0$ and $\bar{h}_{\mu}{}^{\mu} = 0$. (64) is a wave equation involving the Laplace-Beltrami operator term $\bar{h}_{\mu\nu;\alpha}{}^{;\alpha}$ for the Spin-2 gravitational perturbation $\bar{h}_{\mu\nu}$ bearing a mass

$$m_g = \sqrt{2\lambda/3} \quad (65)$$

similar to the Klein-Gordon equation $(\square - m^2)\varphi = 0$ for a Spin-0 scalar field $\varphi$ in flat Minkowski space. Sect. A.3 below demonstrates that $\bar{h}_{\mu\nu;\alpha}{}^{;\alpha} \to \square\bar{h}_{\mu\nu}$ in (64) for the limit $r\to 0$. From (64) and Sect. A.3 then

$$(\square - m_g^2)\bar{h}_{\mu\nu} = -2\kappa\delta T_{\mu\nu} \quad (66)$$

in the locally flat-space limit $r \ll 1$.[22]

By rescaling $\bar{h}$ as $\bar{h}_2 \to \frac{1}{2}\bar{h}_1$ in (59) and (64), (65) becomes

$$m_g = \sqrt{\lambda/3} \quad . \quad (67)$$

which is the surface gravity $\kappa_C = m_g$ of the cosmological event horizon identified by G-H [24].

### A.3 Locally Flat Limit of Wave Equation (64)

It is necessary to demonstrate that hidden $\lambda$-terms arising from $\bar{h}_{\mu\nu;\alpha}{}^{;\alpha}$ in (64) do not cancel the mass term in (64)-(67) when $r\to 0$ and $\bar{h}_{\mu\nu;\alpha}{}^{;\alpha} \to \bar{h}_{\mu\nu,\alpha}{}^{,\alpha} = \square\bar{h}_{\mu\nu}$, the d'Alembertian in a locally flat region of dS studied in Sect. A.2. $\lambda$-terms appear but cancel out as shown below.

To simplify calculations, note that the $r^2 d\Omega^2$ in (3) is of second-order in $r$ and negligible as $r \to 0$. Thus the focus is on $c(r)$ (with $M=0$) in (3) appearing in the diagonal of $\eta_{\mu\nu}$ and its inverse $\eta^{\mu\nu}$. Hence, $\eta_{00} = -c$ and $\eta^{00} = -c^{-1}$, while $\eta_{11} = c^{-1}$ and $\eta^{11} = c$. Also, note that $c(r) \to 1$ and $c(r)^{-1} \to 1$ as $r \to 0$.

Introducing the Christoffel symbol $\Gamma^{\gamma}_{\alpha\beta}$, we can write

$$\bar{h}_{\mu\nu;\alpha}{}^{;\alpha} = g^{\alpha\beta}\bar{h}_{\mu\nu;\alpha;\beta} = g^{\alpha\beta}[\bar{h}_{\mu\nu,\alpha;\beta} - (\Gamma^{\varepsilon}_{\alpha\mu}\bar{h}_{\varepsilon\nu})_{;\beta} - (\Gamma^{\varepsilon}_{\alpha\nu}\bar{h}_{\mu\varepsilon})_{;\beta}] . (68)$$

Define

$$\bar{h}_{\mu\nu;\alpha}{}^{;\alpha} = \square\bar{h}_{\mu\nu} + A_{\mu\nu} + B_{\mu\nu} + C_{\mu\nu} \quad , \quad (69)$$

where

$$\square\bar{h}_{\mu\nu} = \bar{h}_{\mu\nu,\alpha}{}^{,\alpha} \quad (70)$$

$$A_{\mu\nu} = -\Gamma^{\varepsilon}_{\beta\mu}\bar{h}_{\varepsilon\nu}{}^{,\beta} - \Gamma^{\varepsilon}_{\beta\nu}\bar{h}_{\mu\varepsilon}{}^{,\beta} - \Gamma^{\varepsilon}_{\beta\alpha}\bar{h}_{\mu\nu,\varepsilon}\eta^{\alpha\beta}$$
$$-\Gamma^{\varepsilon}_{\alpha\mu}\bar{h}_{\varepsilon\nu}{}^{,\alpha} - \Gamma^{\varepsilon}_{\alpha\nu}\bar{h}_{\mu\varepsilon}{}^{,\alpha} \quad (71)$$

$$B_{\mu\nu} = -(\Gamma^{\varepsilon}_{\alpha\mu})^{\alpha}\bar{h}_{\varepsilon\nu} - (\Gamma^{\varepsilon}_{\alpha\nu})^{\alpha}\bar{h}_{\mu\varepsilon} \quad (72)$$

$$C_{\mu\nu} = -\eta^{\alpha\beta}[(\Gamma^{\varepsilon}_{\beta\delta}\Gamma^{\delta}_{\alpha\mu} - \Gamma^{\delta}_{\beta\alpha}\Gamma^{\varepsilon}_{\delta\mu} - \Gamma^{\delta}_{\beta\mu}\Gamma^{\varepsilon}_{\alpha\delta})\bar{h}_{\varepsilon\nu}$$
$$-\Gamma^{\delta}_{\beta\varepsilon}\Gamma^{\varepsilon}_{\alpha\mu}\bar{h}_{\delta\nu} - \Gamma^{\delta}_{\beta\nu}\Gamma^{\varepsilon}_{\alpha\mu}\bar{h}_{\varepsilon\delta}$$
$$+(\Gamma^{\varepsilon}_{\beta\delta}\Gamma^{\delta}_{\alpha\nu} - \Gamma^{\delta}_{\beta\alpha}\Gamma^{\varepsilon}_{\delta\nu} - \Gamma^{\delta}_{\beta\nu}\Gamma^{\varepsilon}_{\alpha\delta})\bar{h}_{\mu\varepsilon}$$
$$-\Gamma^{\delta}_{\beta\mu}\Gamma^{\varepsilon}_{\alpha\nu}\bar{h}_{\delta\varepsilon} - \Gamma^{\delta}_{\beta\varepsilon}\Gamma^{\varepsilon}_{\alpha\nu}\bar{h}_{\mu\delta}] \quad . \quad (73)$$

$B_{\mu\nu}$ is the term of interest. $A_{\mu\nu}$ and $C_{\mu\nu}$ contain terms of second order, or terms that vanish in locally flat space ($r \ll 1$). Furthermore, only the first-order second derivatives in $B_{\mu\nu}$ remain as $r \to 0$. These terms are

$$B^*_{\alpha\mu\nu}{}^{\alpha} = -\frac{1}{2}\eta^{\varepsilon\gamma}[(\eta_{\alpha\gamma,\mu}{}^{,\alpha} + \eta_{\mu\gamma,\alpha}{}^{,\alpha} - \eta_{\alpha\mu,\gamma}{}^{,\alpha})\bar{h}_{\varepsilon\nu}$$
$$+(\eta_{\alpha\gamma,\nu}{}^{,\alpha} + \eta_{\nu\gamma,\alpha}{}^{,\alpha} - \eta_{\alpha\nu,\gamma}{}^{,\alpha})\bar{h}_{\mu\varepsilon}] \quad (74)$$

which can be defined as

$$B^*_{\alpha\mu\nu}{}^{\alpha} = F_{\mu\nu} + G_{\mu\nu} + H_{\mu\nu} \quad , \quad (75)$$

where

---

[22] The symbol $\square$ in this Appendix refers specifically to an approximately flat space ($r \ll 1$) in (66), as opposed to its more general curved-space meaning in the text used in (24) and (40)-(47). See the close of Sect. A.3.

$$F_{\mu\nu} = -\frac{1}{2}\eta^{\varepsilon\gamma}\left[\,(\Box\eta_{\mu\gamma})\bar{h}_{\varepsilon\nu} + (\Box\eta_{\nu\gamma})\bar{h}_{\mu\varepsilon}\,\right] \quad (76)$$

$$G_{\mu\nu} = -\frac{1}{2}\eta^{\varepsilon\gamma}\left[\,\eta_{\alpha\gamma,\mu}{}^{,\alpha}\bar{h}_{\varepsilon\nu} + \eta_{\alpha\gamma,\nu}{}^{,\alpha}\bar{h}_{\mu\varepsilon}\,\right] \quad (77)$$

$$H_{\mu\nu} = +\frac{1}{2}\eta^{\varepsilon\gamma}\left[\,\eta_{\alpha\mu,\gamma}{}^{,\alpha}\bar{h}_{\varepsilon\nu} + \eta_{\alpha\nu,\gamma}{}^{,\alpha}\bar{h}_{\mu\varepsilon}\,\right] \quad . \quad (78)$$

In this approximation, $\Box = -\partial_t^2 + \nabla^2 \to \nabla^2$. Also $\Box\eta_{00} \to \nabla^2\eta_{00} = +\frac{2}{3}\lambda$ and $\Box\eta_{11} \to \nabla^2\eta_{11} = +\frac{2}{3}\lambda$.

We find that

$$F_{\mu\nu} = -\frac{1}{2}\eta^{00}\left[\,(\Box\eta_{\mu0})\bar{h}_{0\nu} + (\Box\eta_{\nu0})\bar{h}_{\mu0}\,\right]$$
$$-\frac{1}{2}\eta^{11}\left[\,(\Box\eta_{\mu1})\bar{h}_{1\nu} + (\Box\eta_{\nu1})\bar{h}_{\mu1}\,\right] \quad (79)$$

whereby (all other terms do not contribute)

$$F_{00} = -\eta^{00}\left[\,(\Box\eta_{00})\bar{h}_{00}\,\right] = +\frac{2}{3}\lambda\bar{h}_{00} \quad (80)$$

$$F_{11} = -\eta^{11}\left[\,(\Box\eta_{11})\bar{h}_{11}\,\right] = -\frac{2}{3}\lambda\bar{h}_{11} \quad . \quad (81)$$

Next

$$G_{\mu\nu} = -\frac{1}{2}\eta^{11}\left[\,\eta_{11,\mu}{}^{,1}\bar{h}_{1\nu} + \eta_{11,\nu}{}^{,1}\bar{h}_{\mu1}\,\right] \quad (82)$$

whereby (all other terms do not contribute)

$$G_{01} = -\frac{1}{3}\lambda\bar{h}_{01}\,; \quad G_{10} = -\frac{1}{3}\lambda\bar{h}_{10}\,; \quad G_{11} = -\frac{2}{3}\lambda\bar{h}_{11}\,. \quad (83)$$

And lastly,

$$H_{\mu\nu} = \frac{1}{2}\eta^{11}\left[\,\eta_{\alpha\mu,1}{}^{,\alpha}\bar{h}_{1\nu} + \eta_{\alpha\nu,1}{}^{,\alpha}\bar{h}_{\mu1}\,\right] \quad , \quad (84)$$

whereby

$$H_{00} = 0\,; \quad H_{11} = \frac{2}{3}\lambda\bar{h}_{11}\,; \quad H_{01} = \frac{1}{3}\lambda\bar{h}_{01}\,; \quad H_{10} = \frac{1}{3}\lambda\bar{h}_{10}\,. \quad (85)$$

Summarizing, the two contributing terms to $F_{\mu\nu}$ in (80) and (81) are equal and opposite thereby cancelling in (79). Thus, $F_{\mu\nu}=0$. Similarly, the collective $G_{\mu\nu}$ and $H_{\mu\nu}$ terms in (83) and (85) cancel one another, giving $G_{\mu\nu} + H_{\mu\nu} = 0$. Hence $B^*_{\alpha\mu\nu}{}^{\alpha} = B_{\mu\nu} \equiv 0$ in (75) and (72). Therefore $\bar{h}_{\mu\nu;\alpha}{}^{;\alpha} \to \bar{h}_{\mu\nu,\alpha}{}^{,\alpha} = \Box\bar{h}_{\mu\nu}$ in the locally flat limit of (64).

## A.4 Problems with the Pauli-Fierz (P-F) Lagrangian, Averting the vDVZ Discontinuity, and Unitarity

The traditional method for introducing a graviton mass in Spin-2 QG is using the P-F Lagrangian £$_{PF}$ because it does not introduce ghosts. Its Spin-0 helicity can survive in the massless limit, also leading to a JFBD scalar-tensor theory of gravitation [89] as used here.

*Problems with the P-F Lagrangian.* Unfortunately, P-F was originally done on a flat background $\eta_{\mu\nu}$ in (7) which possibly violates the principle of compatible asymptotic states discussed earlier in Sect. 2.1. Therefore, it needs to be re-analyzed. Secondly, it has not seemed to reproduce EG when $m_{PF} \to 0$ (vDVZ below). Thirdly, P-F ignored $\lambda$ ($\lambda$=0). The work-around for this oversight is to conduct the P-F method with $\eta_{\mu\nu}$ representing a de Sitter space. This has been done by Higuchi [95] who obtains (66) above when one assumes $m_{PF} = 0$, a point that reinforces the derivation and conclusions in Sect. A.2.

*vDVZ Discontinuity.* The subject of finite-range gravitation resulted historically in what is known as the vDVZ discontinuity [96-104,89]. In the linear approximation to EG with $m_{PF} \neq 0$, the zero-mass limit $m_{PF} \to 0$ does not produce the same one-graviton propagator as the $m_{PF} = 0$ case. The consequence of the one-graviton approximation is that giving a nonzero mass $m_g = m_{PF}$ to a graviton results in a bending angle of light near the Sun that is 3/4 that of Einstein's value, and the difference may be measurable [89]. The resolution of this QG dilemma is making $m_g$ small enough and not using perturbative approximations [105]. It has since been found that there is no mass discontinuity in the full nonlinear theory [102,103] and none in supergravity [104]. The one-graviton exchange approximation does not produce the correct result for the full nonlinear QG problem [105].

*Unitarity.* Symmetry ($h_{\mu\nu}=h_{\nu\mu}$) and energy conservation reduce the 16 unknowns in $h_{\mu\nu}$ to six (6). By suppressing the vector gravitons ($f_\mu$=0) four (4) more are eliminated, leaving only two (2) for the supposedly "massless" 2S+1 graviton – except that the cosmological term $\lambda$ has survived in (63) and (64) as a mass term. It is a "transverse-traceless" gauge ( $\bar{h}_{\mu\nu}{}^{;\nu} = 0$ and $\bar{h}_\mu{}^\mu = 0$ ) with all of the other spin admixtures removed. $\lambda$ survives because the background $R=4\lambda$ is a curved de Sitter space and can be viewed as the origin of mass in (65)-(67).

For the massive Spin-2 problem with $m_{PF}$, the same procedure does not give the same result. The Hilbert-type gauge condition $f_\mu$=0 does not produce four additional constraints, but rather reduces the field equations in conjunction with energy conservation to provide a relation between the traces $h$ and $T$. That provides only one more constraint, reducing the variables to five (5) or 2S+1 for the Spin-2 massive graviton.[23]

*Bag exterior, free graviton.* From the discussion in Sect. A.2 and without the use of a P-F mass, the free graviton has a tiny $\lambda_{F-L}$-induced mass $m_g \sim 10^{-33} eV$ (Table 1) in the empty space of the hadron exterior that appears naturally in GR. This obviously has a smooth zero-mass limit $m_g \to 0$ as $\lambda \to 0$, as can be seen in (67) above. It has a range equal to the cosmological event horizon radius $r_{EH} = \kappa_C^{-1}$ (Sect. 4). Such a small mass makes $m_g$ immeasurable.

---

[23] Only for a massless particle like the photon must masslessness be invoked, else QFT makes no sense – and hence gauge invariance. However, there is no experimental evidence that the graviton is massless. Hence, gauge invariance for a graviton is an open issue – particularly since EG is not a gauge theory.

*Bag interior, confined graviton*. In the interior case, there is a significant increase in $\lambda=\lambda_{Bag}$ due to the bag constant $B$ as seen in Table 1. The vector gravitons are coupled to the gluon fields via $f_\mu \neq 0$ in (53), relaxing that gauge condition. The Spin-0 graviton component is coupled to the $\sigma$-field via the trace $T^*$ in (45).

There is need to analyze strong interaction physics within hadrons (Figure 2). That may possibly utilize the Pauli-Fierz mass $m_{PF}$ in massive Spin-2 dynamics involving tensor mesons [89] because they can be present. Such calculations are far beyond the scope of this study.

The extent to which $m_{PF}$ plays a part will have to be determined by experiment and a better understanding of strong interaction physics and QCD confinement. Should this type of analysis prove necessary, it can readily proceed from the work of Higuchi [95].